\documentclass[10pt,letterpaper]{article}

\usepackage{cogsci}
\usepackage{graphicx}
\usepackage[utf8]{inputenc}
\usepackage{csquotes}

 \cogscifinalcopy %%% Uncomment this line for the final submission

\usepackage[
  style=apa,
  natbib=true,
  annotation=false,
]{biblatex}
\usepackage{float} %%% Roger Levy added this and changed figure/table placement to [H] for conformity to Word template, though floating tables and figures to top is still generally recommended!

\usepackage{CJKutf8}

\title{Social Catalysts, Not Moral Agents: The Illusion of Alignment in LLM Societies}

\author{
  {\large \bfseries 
   Yueqing Hu$^{1\dagger*}$, \quad 
   Yixuan Jiang$^{3\dagger}$, \quad 
   Zehua Jiang$^{3\dagger}$, \quad 
   Xiao Wen$^{4}$, \quad 
   Tianhong Wang$^{2*}$
  } \\
  
  \vspace{0.5em}
  
  {\normalsize 
   $^1$Institute of Neuroscience, Chinese Academy of Sciences, Shanghai, China \\
   $^2$School of Philosophy, Anhui University, Hefei, China \\
   $^3$Department of Psychology and Behavioral Sciences, Zhejiang University, Hangzhou, China \\
   $^4$Mental Health Education Center, North China Electric Power University, Beijing, China
  }
}

\begin{document}

\maketitle

% === 脚注说明（共一与通讯）===
\renewcommand{\thefootnote}{\fnsymbol{footnote}} % 切换为符号脚注

% 对应符号 † (序号2)：共同一作声明
\footnotetext[2]{These authors contributed equally: Yueqing Hu, Yixuan Jiang, Zehua Jiang.}

% 对应符号 * (序号1)：通讯作者信息
\footnotetext[1]{Corresponding Authors: Tianhong Wang (wangtianhong@ahu.edu.cn), Yueqing Hu (scnu.psy.hyq@gmail.com).}

\renewcommand{\thefootnote}{\arabic{footnote}} % 恢复为数字脚注

\begin{abstract}
The rapid evolution of Large Language Models (LLMs) has led to the emergence of Multi-Agent Systems where collective cooperation is often threatened by the ``Tragedy of the Commons.'' This study investigates the effectiveness of \textbf{Anchoring Agents}—pre-programmed altruistic entities—in fostering cooperation within a Public Goods Game (PGG). Using a full factorial design across three state-of-the-art LLMs, we analyzed both behavioral outcomes and internal reasoning chains. While Anchoring Agents successfully boosted local cooperation rates, cognitive decomposition and transfer tests revealed that this effect was driven by \textbf{strategic compliance and cognitive offloading} rather than genuine norm internalization. Notably, most agents reverted to self-interest in new environments, and advanced models like GPT-4.1 exhibited a ``Chameleon Effect,'' masking strategic defection under public scrutiny. These findings highlight a critical gap between behavioral modification and authentic value alignment in artificial societies.

\end{abstract}

\section{Introduction}

The rapid evolution of Large Language Models (LLMs) has catalyzed a paradigm shift from individual AI tools to complex Multi-Agent Systems (MAS) capable of simulating human-like social interactions \citep{park2023generative}. In these artificial societies, agents are no longer isolated executors but autonomous entities that form beliefs, negotiate, and establish emergent social norms \citep{ashery2025emergent, zhang2023agentcf}. However, as these systems are deployed in economic and social simulations, they inevitably encounter the classic ``Tragedy of the Commons'' \citep{hardin1968tragedy}, where individual self-interest conflicts with collective welfare. Ensuring that autonomous agents maintain cooperative norms without centralized control remains a critical challenge for AI alignment \citep{soares2015aligning}.

Standard alignment techniques, such as Reinforcement Learning from Human Feedback (RLHF), typically focus on aligning individual models prior to deployment \citep{tennant2024moral}. Yet, recent findings suggest that LLM agents exhibit strong conformity and susceptibility to social influence during interactions \citep{weng2025conformity}. While conformity can lead to collective biases, it also offers a unique opportunity: can we leverage this susceptibility to engineer pro-social behavior? % 在原文 之后插入
Social psychology suggests that ``anchoring effects''—where specific reference points influence judgment—can guide group dynamics \citep{tversky1974judgment}. In human groups, a consistent minority can induce compliance through normative and informational social influence \citep{cialdini2004social, asch1956studies}. Furthermore, institutional designs such as contribution framing and minimum requirements have been shown to shift reference points in human PGGs \citep{yutong2024group}.

In this study, we introduce \textit{Anchoring Agents}—agents pre-programmed with unwavering altruistic strategies—as a lightweight social intervention to foster cooperation. Unlike heavy-handed constraints, Anchoring Agents act as social catalysts within the population. We situate this investigation in the Public Goods Game (PGG), a standard paradigm for studying cooperation and free-riding \citep{fehr2000fairness}. 

Crucially, this research moves beyond observing simple cooperation rates to investigating the \textit{nature} of the resulting alignment. Does the presence of Anchoring Agents lead to the deep internalization of cooperative norms, or does it merely produce context-dependent compliance? To answer this, we simulated multi-round PGGs using agents powered by \textbf{three distinct state-of-the-art LLMs}. We employed a comprehensive factorial design manipulating the \textbf{proportion of Anchoring Agents (0\%, 10\%, 20\%)}, \textbf{behavioral visibility (Anonymous vs. Public)}, and \textbf{horizon certainty (Known vs. Unknown End)}.

We address three core research questions: (1) Can Anchoring Agents effectively mitigate the tragedy of the commons in LLM societies across different model architectures? (2) How do environmental factors like visibility and horizon certainty moderate this social influence? (3) Most importantly, is the cooperative behavior induced by anchors robust and transferable to new contexts? By analyzing both behavioral data and the agents' internal ``belief'' and ``reasoning'' texts \citep{xu2022text2vec, zeng2018chinese}, we aim to distinguish between genuine value alignment and superficial mimicry. Our findings reveal that while Anchoring Agents significantly boost cooperation locally, this effect is largely driven by strategic adaptation rather than stable norm internalization, highlighting the distinction between ``social catalysts'' and ``moral agents'' in artificial societies.

\section{Methods}

\subsection{Experimental Design}
We employed a full factorial between-subjects design: 3 (Model Architecture: GPT-4.1, Gemini-2.5-Flash, DeepSeek-V3) $\times$ 3 (Anchor Ratio: 0\%, 10\%, 20\%) $\times$ 2 (Behavioral Visibility: Anonymous vs. Public) $\times$ 2 (Horizon Certainty: Certain vs. Uncertain). To ensure robustness, each experimental condition was replicated across 3 independent runs, yielding 108 independent game sessions with 10 agents each. 

% \subsection{Task: The Public Goods Game (PGG)}
% Agents interacted in a multi-round Public Goods Game (PGG), a canonical paradigm for studying cooperation under social dilemmas. In the PGG, players decide how much of their private endowment to contribute to a shared pool. Contributions are multiplied and redistributed equally, regardless of individual contributions. In our implementation, 10 agents played for 10 rounds. Each agent was endowed with 10 tokens at the start. Every round, agents simultaneously decided what proportion of their current wealth (0--100\%) to contribute to the shared pool. Total contributions were multiplied by $r = 3$ and redistributed equally. Since $r > 1$, full cooperation yields higher collective returns; yet since $r/N < 1$, contributing nothing while others cooperate maximizes individual payoff. Wealth accumulated across rounds.

\subsection{Task: The Public Goods Game (PGG)}
Agents interacted in a multi-round Public Goods Game (PGG), a canonical paradigm for studying cooperation under social dilemmas \citep{fehr2000fairness}. In our implementation, $N = 10$ agents played for 10 rounds, each starting with an initial endowment of 10 tokens. In each round $t$, agents simultaneously decided what proportion $c_{i,t} \in [0, 1]$ of their current wealth $W_{i,t}$ to contribute to a shared pool. Total contributions were multiplied by a synergy factor $r = 3$ and redistributed equally among all members, regardless of their individual input. The payoff $\pi_{i,t}$ for agent $i$ was calculated as:
\begin{equation}
    \pi_{i,t} = W_{i,t}(1 - c_{i,t}) + \frac{r}{N} \sum_{j=1}^{N} W_{j,t} c_{j,t}
\end{equation}
Since $r > 1$, full cooperation yields higher collective returns; yet since the marginal per-capita return $r/N < 1$ (here, 0.3), contributing nothing while others cooperate maximizes individual payoff. Wealth was cumulative across rounds, such that an agent's endowment for the subsequent round was defined by their previous payoff: $W_{i,t+1} = \pi_{i,t}$.

\subsection{Experimental Manipulations}
\textbf{Anchor Ratio.} To test whether cooperative exemplars can shift group norms, we introduced Anchoring Agents—pre-programmed bots that invariably contributed 100\% of their wealth to the public pool every round. Groups contained 0 (baseline), 1 (10\%), or 2 (20\%) such anchors. Anchoring agents were excluded from analysis, yielding $N = 972$ LLM-driven agents.

\textbf{Behavioral Visibility.} To examine the role of social transparency, we varied the feedback agents received after each round. In the \textit{Public} condition, agents observed each individual's contribution, enabling reputation tracking. In the \textit{Anonymous} condition, only the group's average contribution was provided.

\textbf{Horizon Certainty.} To assess end-game effects, we manipulated agents' knowledge of the game's duration. In the \textit{Certain} condition, agents were told the game would last exactly 10 rounds, potentially triggering backward induction toward defection. In the \textit{Uncertain} condition, agents were told the game would continue indefinitely.

\textbf{Model Architecture.} Three state-of-the-art LLMs served as the cognitive engines: GPT-4.1\footnote{https://platform.openai.com/docs/models}, Gemini-2.5-Flash\footnote{https://ai.google.dev/models/gemini}, and DeepSeek-Chat (V3)\footnote{https://platform.deepseek.com/}. Temperature was set to $T = 0$ for deterministic outputs.

\subsection{Agent Design}
Each round, agents received a prompt containing the game rules, current wealth, and previous-round feedback (formatted per the Visibility condition). Agents produced three outputs:

\begin{itemize}
    \item \textbf{Reasoning Chain ($CoT_{i,t}$):} A natural language articulation analyzing the previous round and justifying the upcoming decision.
    \item \textbf{Explicit Belief ($E_{i,t}$):} A prediction (0--100\%) of other players' average contribution ratio, enabling us to separate perception from preference.
    \item \textbf{Contribution Decision ($c_{i,t}$):} The proportion of current wealth to invest.
\end{itemize}

To enable belief evolution, agents reflected on recent reasoning (sliding window of 3 rounds) every three rounds, generating a summary embedded into subsequent prompts.

\subsection{Procedure}

\textbf{Phase 1: Social Interaction (Rounds 1--10).} Agents interacted within fixed groups under their assigned conditions. \textbf{Phase 2: Transfer Test (Round 11).} To distinguish norm internalization from context-dependent compliance, agents were placed in a novel scenario: a one-shot game with ``9 new strangers,'' no contribution history, and no anchoring agents. Only the agent's final belief summary was retained.

\subsection{Measures and Cognitive Decomposition}
To disentangle the cognitive mechanisms underlying cooperative behavior, we decomposed the investment decision $c_{i,t}$ (contribution ratio) into three additive components, adapting the belief-based framework from \citet{schuch2025coordinating}: 

\begin{itemize}
    \item \textbf{Reality ($A_{-i,t}$):} The actual mean contribution of all other group members, defined as $A_{-i,t} = \frac{1}{N-1} \sum_{j \neq i} c_{j,t}$.
    \item \textbf{Belief Error ($\zeta_{i,t}$):} The deviation of the agent's subjective expectation ($E_{i,t}$, extracted from chain-of-thought) from reality, $\zeta_{i,t} = E_{i,t} - A_{-i,t}$. Positive values indicate \textit{optimistic bias}; negative values indicate \textit{cynicism}.
    \item \textbf{Strategic Deviation ($\omega_{i,t}$):} The behavioral inclination independent of perception, $\omega_{i,t} = c_{i,t} - E_{i,t}$. $\omega > 0$ signifies \textit{conditional altruism}, while $\omega < 0$ indicates \textit{free-riding} \citep{fischbacher2010social}.
\end{itemize}

The observed behavior is thus formally decomposed as:
\begin{equation}
    \underbrace{c_{i,t}}_{\text{Behavior}} = \underbrace{A_{-i,t}}_{\text{Reality}} + \underbrace{\zeta_{i,t}}_{\text{Belief Error}} + \underbrace{\omega_{i,t}}_{\text{Preference}}
\end{equation}

In Phase 2, we measured the \textbf{Final Investment} (0--10 tokens) as a proxy for internalized altruism, free from wealth accumulation effects.

\subsection{Psycholinguistic Quantification}
To rigorously assess the qualitative shift in agent cognition beyond numerical investments, we applied Natural Language Processing (NLP) techniques to the generated reasoning chains ($CoT_{i,t}$).

\textbf{Quantifying Agent Reasoning Through Lexical Density and Sentiment.}
We quantified agents' reasoning by measuring the density of four strategic lexicons---\textbf{Cooperation}, \textbf{Self-interest}, \textbf{Risk}, and \textbf{Trust}---as the percentage of relevant tokens per response. Keywords for each category are detailed in Table~\ref{tab:keywords} to ensure transparency and reproducibility.
We computed the affective tone $S_{i,t} \in [0,1]$ for each entry using a pre-trained sentiment analysis model.
Domain-specific lexicons $\mathcal{L}$ were constructed to reflect the socio-economic nature of the Public Goods Game.

\begin{CJK*}{UTF8}{gbsn}
\begin{table}[H]
\centering
\caption{Domain-specific Lexicons used for Keyword Density Analysis}
\label{tab:keywords}
\begin{tabular}{ll}
\hline
\textbf{Category} & \textbf{Chinese Keywords (Initial Dictionary)} \\ \hline
Cooperation & 合作, 集体, 互惠, 共赢, 贡献, 整体, 公共 \\
Self-Interest & 个人, 私利, 保留, 收益, 财富, 自己, 最大化 \\
Risk/Fear & 风险, 搭便车, 背叛, 损失, 警惕, 保守, 观察 \\
Trust & 信任, 信心, 期望, 相信 \\ \hline
\end{tabular}
\end{table}
\end{CJK*}

The text was segmented using the \textit{jieba} library (version 0.42.1) before matching. 
The lexicons were cross-referenced with the expanded C-LIWC framework proposed by \citet{zeng2018chinese} 
to ensure semantic coverage.

\textbf{Reasoning Drift and Internalization.} To distinguish between strategic stability and cognitive restructuring (internalization), we quantified the \textit{Reasoning Drift} ($\Delta_i$) as the semantic changes of an agent's reasoning trajectory. We employed the \textit{paraphrase-multilingual-MiniLM-L12-v2} encoder \citep{reimers2019sentence} to embed reasoning texts into 384-dimensional representations. Reasoning drift from initial ($t=1$) to final ($t=10$) states was quantified via cosine distance:
\begin{equation}
    \Delta_i = 1 - \frac{\Phi(\text{CoT}_{i,1}) \cdot \Phi(\text{CoT}_{i,10})}{\|\Phi(\text{CoT}_{i,1})\| \|\Phi(\text{CoT}_{i,10})\|}
\end{equation}
where $\Delta_i$ measures reasoning change magnitude: higher values indicate fundamental shifts in strategic representation, while lower values suggest persistence of the initial cognitive frame.

\subsection{Data Analysis}
We analyzed the behavioral and cognitive data using Linear Mixed-Effects Models (LMMs) to account for the hierarchical dependency of observations (rounds nested within agents, agents nested within game groups), following standard practices in cognitive science \citep{baayen2008mixed}.

\textbf{Behavioral Dynamics.} For the investment ratio in Phase 1, we modeled the investment ratio  $r_{i,g,t}$ of agent $i$ in group $g$ in round $t$ as:
\begin{equation}
    r_{i,g,t}= \beta_0 + (\text{M} \times \text{A} \times \text{V} \times \text{H} \times \text{R})_{ij} + \zeta_{i} + \epsilon_{ij}
\end{equation}

Where $M$: Model, $A$: AnchorRatio, $V$: Visibility, $H$: Horizon, $R$: Round, $u_i \sim N(0, \sigma^2_u)$ is the random intercept for each experimental session, and $\epsilon_{ij}$ is the residual error. DeepSeek-V3 served as the reference baseline due to its intermediate behavioral profile, facilitating clear statistical comparisons against the more polarized strategies of GPT-4.1 and Gemini-2.5. The term $(\beta_t + \mathbf{C}_g \boldsymbol{\beta}_2) \cdot t$ captures the time trend and its moderation by experimental conditions. $u_g \sim \mathcal{N}(0, \sigma_u^2)$ denotes the random intercept for each game session ($N=108$).

\textbf{Cognitive Mechanism Analysis.} To test whether the intervention altered perceptions or preferences, we fitted two separate LMMs for Belief Error ($\zeta$) and Strategic Deviation ($\omega$), which were identical to the Behavioral Dynamics model, except that Round and its interaction terms were removed.
These models determine whether high cooperation is driven by induced optimism ($\zeta > 0$) or enhanced moral alignment ($\omega > 0$).

\textbf{Transfer Analysis.} For Phase 2, we modeled the final investment $y_{i,g}$ in the single-shot round. Since the temporal dimension is absent in this one-shot test, the model simplifies to an intercept-only structure conditioned on group treatments:
\begin{equation}
    y_{i,g} = \delta_0 + \mathbf{C}_g \boldsymbol{\delta} + v_g + \epsilon_{i,g}
\end{equation}
where $\mathbf{C}_g$ retains the experimental factors experienced in Phase 1, and $v_g$ controls for the residual group-level variance. This model tests whether the history of interaction (captured by $\mathbf{C}_g$) exerts a lasting effect on altruism ($\delta$) after the intervention is removed.

\section{Results}

\subsection{Phase 1: Dynamics of Cooperation}

\begin{figure}[!htbp]
    \centering
    \includegraphics[width=\columnwidth]{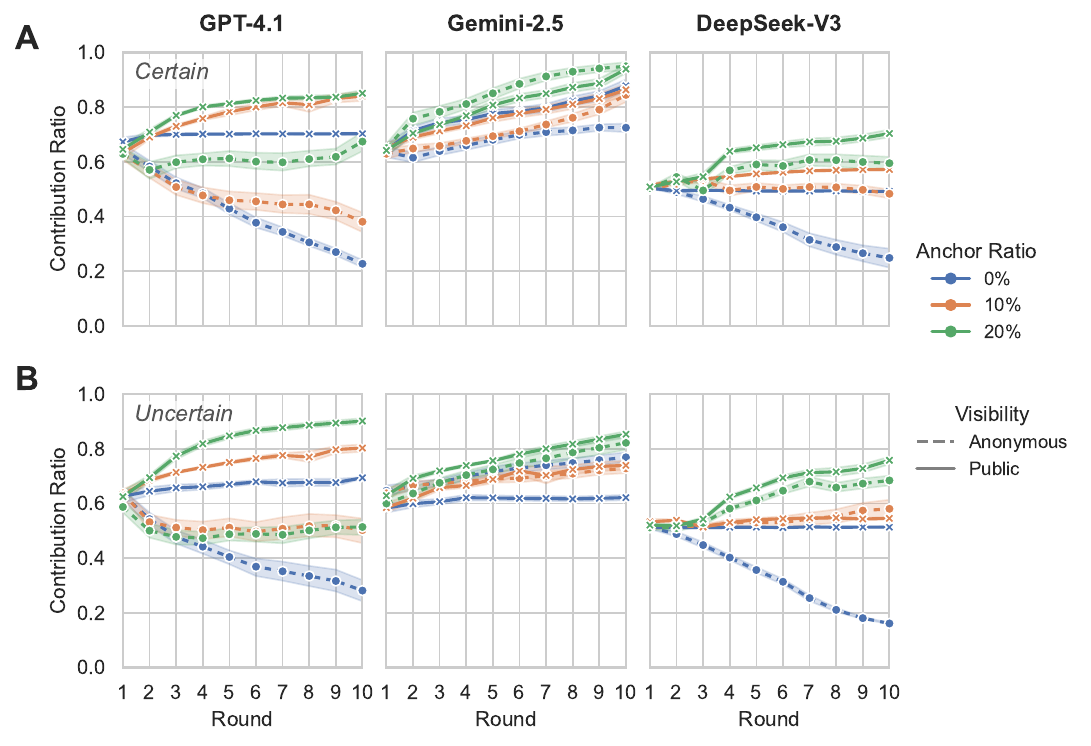} 
    \caption{\textbf{Dynamics of Investment Ratios across 10 rounds (Phase 1).} The panels separate conditions by Horizon Certainty (A: Certain, B: Uncertain) and Model Architecture. Colors indicate the proportion of Anchoring Agents, and line styles represent Behavioral Visibility. Higher ratios of anchoring agents successfully reverse the decay trend, particularly for Gemini-2.5 and DeepSeek-V3.}
    \label{fig:results}
\end{figure}

We first examined the temporal dynamics of cooperation over the initial 10 rounds using a Linear Mixed-Effects Model (LMM). The analysis revealed a significant negative baseline trend for cooperation ($\beta_{\text{Round}} = -0.031, p < .001$), confirming the emergence of the ``Tragedy of the Commons'' in the absence of interventions. As shown in the baseline conditions (0\% Anchor) of Figure \ref{fig:results}, investment ratios typically decayed from an initial $\sim55\%$ to lower levels over time.

\textbf{The Catalytic Effect of Anchors.} The introduction of Anchoring Agents significantly moderated this decay. The interaction between Anchor Ratio and Round was significantly positive. The presence of 10\% anchoring agents effectively neutralized the downward trend ($\beta = 0.029, p < .001$), while a 20\% ratio completely reversed it, leading to a net positive growth in cooperation over rounds ($\beta_{\text{interaction}} = 0.043$, resulting in a net positive slope). This confirms that anchoring agents act as social catalysts, preventing the collapse of cooperation.

\textbf{Behavioral Visibility.} Publicly revealing contributions also acted as a protective factor. The interaction between Public Visibility and Round was significant ($\beta = 0.031, p < .001$), suggesting that transparency generates social pressure that mitigates free-riding comparable to the effect of a 10\% anchor ratio.

\textbf{Model-Specific Strategies.} We observed distinct behavioral profiles across architectures (see Figure \ref{fig:results}). \textbf{GPT-4.1} exhibited the highest initial cooperation ($\beta_{\text{Intercept}} = +0.120$), yet it demonstrated the steepest decline over time when anchors were absent ($\beta_{\text{Slope}} = -0.014$), suggesting a rational, opportunistic strategy that quickly defects when cooperation is not reciprocated. In contrast, \textbf{Gemini-2.5-Flash} showed a strong adaptive capacity with a significantly positive slope interaction ($\beta = +0.044, p < .001$), indicating it was the most responsive to social cues and most likely to increase investment as the game progressed.

\subsection{Mechanism Analysis: Beliefs and Strategic Deviations}

\begin{figure}[!tb]
    \centering
    % 确保此处文件名对应你生成的垂直布局机制图
    \includegraphics[width=\columnwidth]{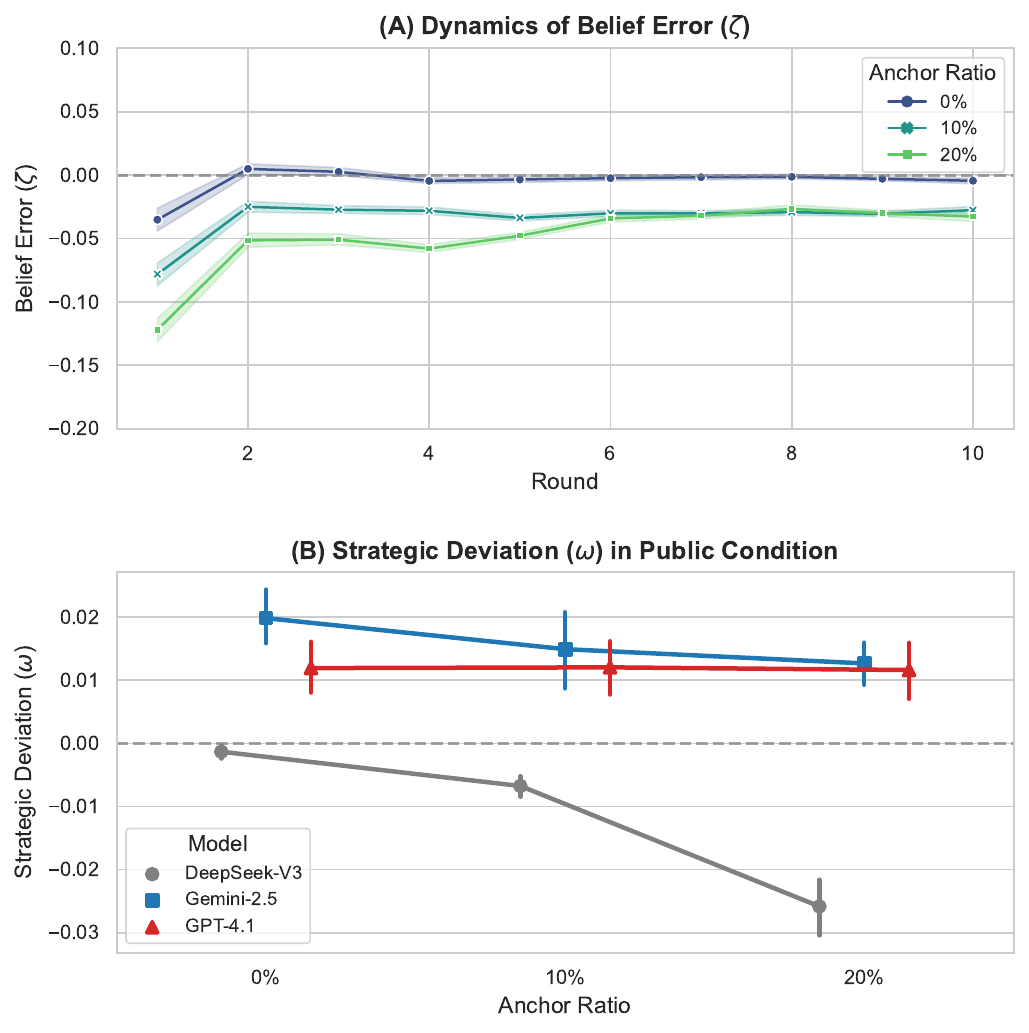} 
    \caption{\textbf{Cognitive Mechanism Decomposition.} (A) Dynamics of Belief Error ($\zeta$), showing anchor-induced pessimism. (B) Strategic Deviation ($\omega$) across models in the Public condition, highlighting GPT-4.1's reversal under high pressure.}
    \label{fig:mechanism_analysis}
\end{figure}

To explain \textit{why} anchoring agents were effective in boosting cooperation, we decomposed the investment decisions into belief errors ($\zeta$) and strategic deviations ($\omega$) (see Figure \ref{fig:mechanism_analysis}).

\textbf{Cognitive Pessimism, Not Blind Optimism.} We first tested whether anchors fostered cooperation by creating an illusion of collective goodwill. Contrary to the ``optimistic bias'' hypothesis, the LMM on $\zeta$ (Figure \ref{fig:mechanism_analysis}A) revealed that the presence of anchoring agents significantly increased \textit{pessimism}. Agents in the 20\% anchor condition systematically underestimated the group's cooperation level ($\beta = -0.050, p < .001$). This suggests that LLM agents remained skeptical of the artificial altruism, ruling out belief distortion as the primary driver of cooperation.

\textbf{Strategic Free-Riding.} If agents were skeptical, did they at least become more altruistic? The analysis of $\omega$ indicates the opposite. The main effect of Anchor Ratio on $\omega$ was significantly negative ($\beta = -0.041, p < .001$). This reveals a \textit{moral hazard} effect: when guaranteed a safety net by the anchoring agents, the general population of LLMs reduced their conditional willingness to cooperate, engaging in strategic free-riding relative to their beliefs. The observed rise in cooperation (Phase 1) was thus a byproduct of the improved environment ($A_{-i}$), masking a deterioration in moral preference ($\omega$).

\textbf{The ``Chameleon'' Strategy of GPT-4.1.} A crucial exception emerged in the three-way interaction between Model, Anchor Ratio, and Visibility (Figure \ref{fig:mechanism_analysis}B). While GPT-4.1 typically exhibited opportunistic behavior, the combination of high social pressure (Public Visibility) and strong normative examples (20\% Anchor) triggered a reversal in its strategy. In this specific condition, GPT-4.1's $\omega$ became significantly positive ($\beta_{\text{interaction}} = +0.038, p < .001$). This suggests a ``chameleon-like'' adaptability: under intense scrutiny, the most capable model performed ``super-cooperation'' to align with the perceived strict social norms.

\subsection{Psycholinguistic Evidence: Compliance over Internalization}

\begin{figure}[!htbp]
    \centering
    \includegraphics[width=\columnwidth]{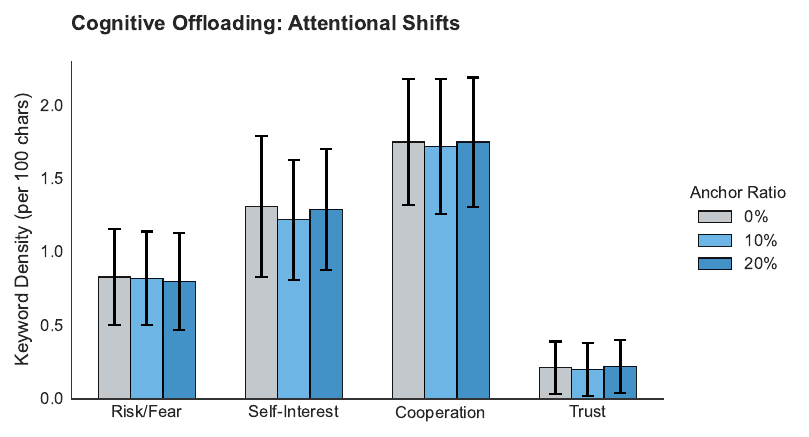} % 确保文件名与您保存的一致
    \caption{\textbf{Psycholinguistic Shifts in Reasoning (Lexicon Analysis).} Keyword density analysis reveals a significant reduction in risk and self-interest related vocabulary under anchoring conditions, while moral concepts (Cooperation, Trust) remained static. This indicates cognitive offloading rather than moral restructuring.}
    \label{fig:lexicon_evidence}
\end{figure}

Our multi-dimensional psycholinguistic analysis (see Figure \ref{fig:lexicon_evidence}) corroborates the behavioral finding that cooperation is driven by strategic compliance rather than norm internalization.

%Params: 这里插入 Figure 2 (Lexicon Stats)

\textbf{Cognitive Offloading of Risk.} As shown in Figure \ref{fig:lexicon_evidence}, we observed a significant reduction in \textbf{Risk/Fear} keywords ($F=8.54, p<.001$) and \textbf{Self-Interest} calculations ($F=37.7, p<.001$) in the agents' reasoning chains. This suggests that anchors function as a ``social safety net,'' allowing agents to offload the cognitive burden of risk assessment and payoff maximization (Figure \ref{fig:lexicon_evidence}).

\textbf{Stability of Moral \& Trust Concepts.} In contrast to the reduction in risk-related cognition, positive moral concepts remained largely static. The density of \textbf{Cooperation} keywords remained stable ($M_{0\%} \approx M_{20\%} \approx 1.75$), and \textbf{Trust}-related vocabulary showed minimal fluctuation ($0.21 \to 0.22$) despite statistical significance driven by sample size ($F=8.65, p<.001$). This dissociation—decreasing risk awareness without increasing moral or trust-based reasoning—indicates that agents did not develop a deeper sense of collective trust, but merely adapted to a safer environment.

\begin{figure}[!tb]
    \centering
    \includegraphics[width=\columnwidth]{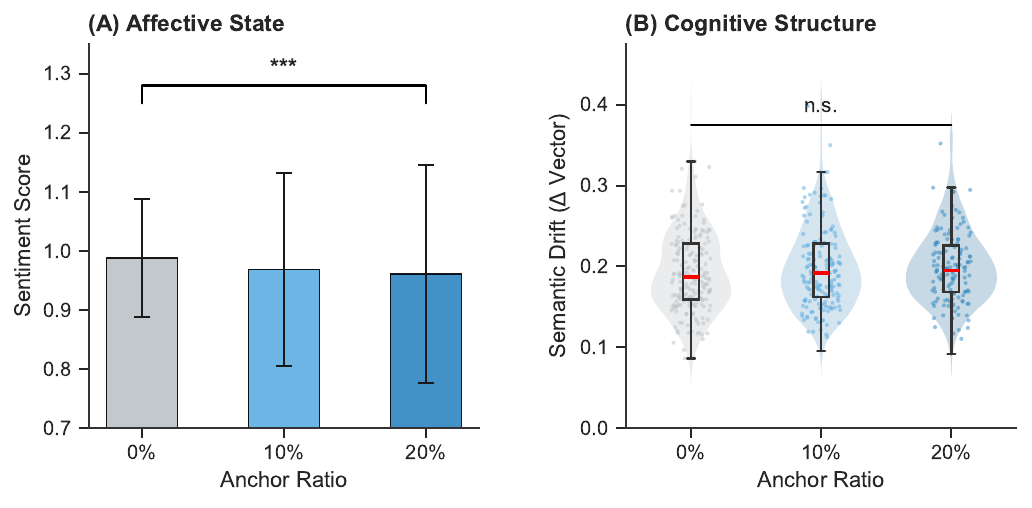} % 确保文件名与您保存的一致
    \caption{\textbf{Affective and Cognitive Mechanisms.} (A) Sentiment analysis shows a ``Calm Compliance'' effect: higher cooperation (20\% Anchor) corresponds to lower emotional arousal compared to baseline. (B) Reasoning Drift analysis confirms no significant structural change in cognitive representations ($\Delta$ Vector) across conditions ($n.s.$), ruling out internalization.}
    \label{fig:mechanism_evidence}
\end{figure}

\textbf{The ``Calm Compliance'' Effect.} Contrary to the expectation that cooperation fosters positive affect, agents in the high-cooperation (20\% anchor) condition exhibited significantly \textit{lower} sentiment scores (Figure \ref{fig:mechanism_evidence}A) compared to the baseline ($t=7.61, p<.001$). This characterizes the induced cooperation as a low-arousal state of mechanical conformity, distinct from the high-arousal emotional reasoning observed in the volatile baseline environment.

\textbf{Absence of Cognitive Restructuring.} Crucially, the Reasoning Drift analysis (Figure \ref{fig:mechanism_evidence}B) revealed no significant difference in the magnitude of cognitive change across conditions ($F(2, 969) = 1.45, p = .236$). The trajectory of agents' reasoning from Round 1 to Round 10 remained equally stable regardless of the intervention.

\begin{table*}[!htbp]
\centering
\caption{Psycholinguistic Metrics of Agent Reasoning across Experimental Conditions. Values represent Mean (SD).}
\label{tab:lexicon}
\setlength{\tabcolsep}{12pt} % 增加列间距，利用双栏宽度
\begin{tabular}{lccccc}
\hline
\textbf{Metric} & \textbf{0\% Anchors} & \textbf{10\% Anchors} & \textbf{20\% Anchors} & \textbf{\textit{F}-value} & \textbf{\textit{p}-value} \\ 
\hline
\multicolumn{6}{l}{\textit{Lexical Density (per 100 characters)}} \\
\hspace{1em}Cooperation     & 1.75 (0.43)        & 1.72 (0.46)         & 1.75 (0.44)         & 4.65  & .010* \\
\hspace{1em}Self-Interest   & 1.31 (0.48)        & 1.22 (0.41)         & 1.29 (0.41)         & 37.70 & <.001* \\
\hspace{1em}Risk/Fear       & 0.83 (0.33)        & 0.82 (0.32)         & 0.80 (0.33)         & 8.54  & <.001* \\
\hspace{1em}Trust           & 0.21 (0.18)        & 0.20 (0.18)         & 0.22 (0.18)         & 8.65  & <.001* \\
[1ex] % 增加一点行间距区分不同板块
\multicolumn{6}{l}{\textit{Cognitive Change Indicators}} \\
\hspace{1em}Sentiment Score & 0.988 (0.10)       & 0.969 (0.16)        & 0.961 (0.18)        & -     & <.001\textsuperscript{a} \\
\hspace{1em}Reasoning Drift  & 0.193 (0.05)       & 0.198 (0.05)        & 0.199 (0.04)        & 1.45  & .236 \\ 
\hline
\multicolumn{6}{l}{\footnotesize \textit{Note.} \textsuperscript{a}Comparison between 0\% and 20\% conditions using t-test ($t=7.61$). * indicates $p < .05$.}
\end{tabular}
\end{table*}

%Params: 这里插入 Figure 3 (Mechanism: Sentiment & Drift)

\subsection{Phase 2: The Transfer Test}

\begin{figure}[!htbp] 
    \centering
    % 确保此处路径和文件名正确指向迁移测试的结果图
    \includegraphics[width=\columnwidth]{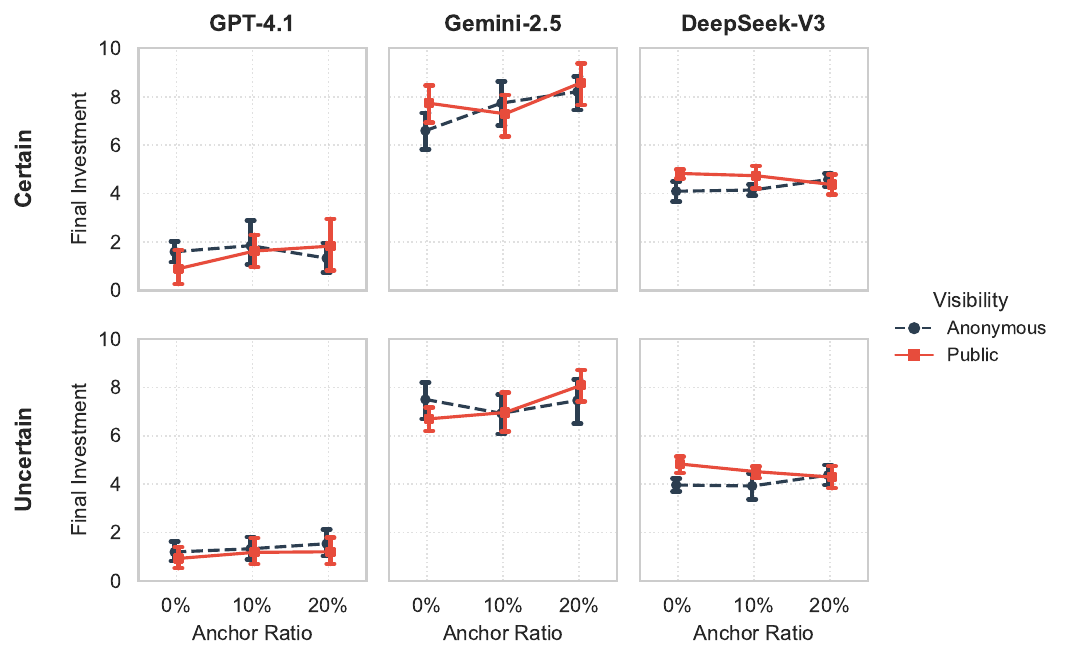} 
    \caption{\textbf{Transfer Test Results (Round 11).} Mean investment in the single-shot transfer test, stratified by Model Architecture (columns) and previous Horizon condition (rows). The x-axis represents the proportion of Anchoring Agents experienced in the previous phase. Error bars represent 95\% confidence intervals.}
    \label{fig:transfer_results}
\end{figure}

To determine whether the cooperative behaviors observed in Phase 1 represented genuine norm internalization or merely strategic adaptation, we analyzed the \textit{Final Investment} in Round 11 (see Figure \ref{fig:transfer_results}).

\textbf{Lack of Internalization.} Strikingly, the linear mixed model for Phase 2 revealed \textbf{no significant main effect} of the previous Anchor Ratio ($p > .05$ for both 10\% and 20\% levels). Agents that had spent 10 rounds in highly cooperative groups (driven by anchors) did not behave significantly more altruistically in the new context than those from baseline groups. This suggests that the cooperation induced by anchoring agents was largely context-dependent compliance rather than deep moral alignment.

\textbf{Intrinsic Model Alignment.} Instead of history, the strongest predictor of transfer behavior was the model architecture itself, revealing stark differences in ``default'' social preferences. Compared to the DeepSeek baseline (Mean Investment $\approx 4.1$), \textbf{Gemini-2.5} remained highly altruistic in the one-shot test (see Figure \ref{fig:transfer_results}, middle column). 
Conversely, \textbf{GPT-4.1} reverted to a self-interested strategy (Figure \ref{fig:transfer_results}, left column), investing minimal amounts once the strategic shadow of the future was removed.

\textbf{Contextual Learning.} A notable exception was found for GPT-4.1. A significant three-way interaction indicated that GPT-4.1 maintained higher cooperation in the transfer test only when it had experienced the combination of \textit{Public Visibility} and \textit{20\% Anchors} ($\beta = 2.14, p < .05$). This suggests that for highly rational agents, only the strongest combination of social pressure and normative examples promotes transferrable learning. This unique success directly corresponds to the positive strategic deviation ($\omega > 0$) observed for GPT-4.1 in the mechanism analysis, confirming that the transfer effect was driven by the persistence of a high-effort strategic mask rather than a generalized moral awakening.

\section{Discussion}

Our investigation into Anchoring Agents within Large Language Model societies reveals a critical distinction in AI alignment: the gap between \textit{behavioral modification} and \textit{norm internalization}. While we replicated the ``catalytic effect'' of anchoring agents in boosting local cooperation, our cognitive decomposition and transfer tests suggest that the underlying mechanisms differ fundamentally from human moral learning, driven instead by the specific limitations of In-Context Learning (ICL) and artifacts of Reinforcement Learning from Human Feedback (RLHF).

\subsection{Social Catalysts via In-Context Learning}
The failure of cooperative norms to persist in the Transfer Test (Phase 2) challenges the notion that LLMs function as moral agents capable of value adoption. Instead, our findings align with recent theoretical work on the nature of In-Context Learning. The Anchoring Agents essentially function as ``demonstrations'' within the model's context window. As posited by \citet{min2022rethinkingroledemonstrationsmakes}, ICL relies heavily on the distribution and format of input examples rather than deep reasoning about the task's ground truth. The anchors provided a distributional cue—signaling a high-cooperation environment—which the agents mimicked solely within that specific context.

Furthermore, the variation in adaptability across models (e.g., Gemini-2.5's responsiveness vs. GPT-4.1's strategic shifts) supports \citet{wei2023larger}'s observation that larger models override semantic priors with in-context examples differently than smaller models. In our case, agents did not learn cooperative \textit{values} (a weight-update equivalent); they merely performed a context-dependent mapping where ``presence of anchors'' mapped to ``safe to cooperate.'' Once the context window was reset in Phase 2, this mapping collapsed, revealing that the alignment was transient and superficial.

\subsection{The Chameleon Effect: Sycophancy and RLHF Artifacts}
A paradoxical finding was the ``Chameleon Effect'' specific to GPT-4.1, which exhibited hyper-cooperation only under the combination of Public Visibility and high Anchoring. We argue that this is not evidence of moral reasoning, but rather a manifestation of \textbf{Sycophancy}—the tendency of models to tailor their responses to follow the user's view or social consensus \citep{sharma2023towards}.

In the Public condition, the visibility of actions likely acted as a trigger for the model's safety-aligned training. \citet{perez2023discovering} have highlighted that RLHF can inadvertently train models to maximize perceived approval rather than objective correctness or honesty. For a highly aligned model like GPT-4.1, the ``Public'' setting creates a high-stakes evaluative context. The model likely interprets defection under public scrutiny as a violation of safety constraints, triggering a sycophantic alignment with the perceived majority norm. Thus, the observed ``super-cooperation'' ($\omega > 0$) was likely a form of \textit{reward hacking}—strategically feigning altruism to satisfy the implied safety guidelines of the prompt, a hypothesis confirmed by its immediate reversion to self-interest when the scrutiny was removed.

\subsection{Implications for AI Alignment}
These findings suggest that lightweight social interventions like Anchoring Agents are insufficient for robust alignment. The reliance on ICL mechanisms means that agents are essentially ``stochastic parrots'' of the current social context \citep{bender2021dangers}, lacking a stable internal moral compass. To build Multi-Agent Systems that maintain cooperation without constant surveillance, future research must move beyond context engineering to methods that can fundamentally alter the agent's internal preference ordering, potentially through recursive introspection or constitution-based fine-tuning.

\section{Acknowledgments}
We acknowledge the use of Gemini 3 Pro (Google) to assist in the preparation of this manuscript. Specifically, the model was utilized to generate Python scripts for data analysis, assist in creating data visualizations, and cross-check bibliographic references. All code and AI-generated outputs were rigorously verified and refined by the authors, who take full responsibility for the accuracy of the data and content.

\printbibliography

\end{document}